\documentclass[12pt,a4paper]{article}
\usepackage[utf8]{inputenc}
\usepackage[T1]{fontenc}
\usepackage{lmodern}
\usepackage{amsmath,amssymb}
\usepackage{graphicx}
\usepackage{caption}
\usepackage{subcaption}
\usepackage{hyperref}
\usepackage{geometry}
\usepackage{setspace}
\usepackage{algorithm}
\usepackage{algpseudocode}
\usepackage{pgfplots}
\usepackage{booktabs}
\usepackage{fancyhdr}
\fancypagestyle{firstfooter}{  
  \fancyhf{}

  \fancyfoot[C]{\scriptsize
  © 2025 IEEE. Personal use of this material is permitted. Permission from IEEE must be obtained for all other uses, in any current or future media.\\
  This is the accepted version of the paper that will appear in the 2025 IEEE 7th International Conference on Blockchain Computing and Applications (BCCA).\\
  Final version will be published at: [ ]}
}

\title{Blockchain Epidemic Consensus for Large-Scale Networks}

\author{
  Siamak Abdi\\
  \small Free University of Bolzano, Italy\\
  \small \texttt{sabdi@unibz.it}
  \and
  Giuseppe Di Fatta\\
  \small Free University of Bolzano, Italy\\
  \small \texttt{giuseppe.difatta@unibz.it}
  \and
  Atta Badii\\
  \small University of Reading, United Kingdom\\
  \small \texttt{atta.badii@reading.ac.uk}
  \and
  Giancarlo Fortino\\
  \small University of Calabria, Italy\\
  \small \texttt{giancarlo.fortino@unical.it}
}

\date{July 2025}

\begin{document}
\maketitle
\thispagestyle{firstfooter}
\begin{abstract}
Blockchain is a distributed ledger technology that has applications in many domains such as cryptocurrency, smart contracts, supply chain management, and many others. Distributed consensus is a fundamental component of blockchain systems that enables secure, precise, and tamper-proof verification of data without relying on central authorities. Existing consensus protocols, nevertheless, suffer from drawbacks, some of which are related to scalability, resource consumption, and fault tolerance.
We introduce Blockchain Epidemic Consensus Protocol (BECP), a novel fully decentralised consensus protocol for blockchain networks at a large scale. BECP follows epidemic communication principles, without fixed roles like validators or leaders, and achieves probabilistic convergence, efficient message dissemination, and tolerance to message delays. We provide an extensive experimental comparison of BECP against classic protocols like PAXOS, RAFT, and PBFT, and newer epidemic-based protocols like Avalanche and Snowman. The findings indicate that BECP provides desirable gains in throughput, consensus latency, and substantial message-passing efficiency compared to existing epidemic-based approaches, validating its usability as an effective and scalable approach for next-generation blockchain systems.
\end{abstract}

\section{Introduction}
Blockchain is a Distributed Ledger Technology (DLT) that structures data into an immutable chain of cryptographically connected blocks, replicated on a network of nodes. The design provides transparency, tamper-evidence, and decentralised trust, without requiring central authorities. Blockchain's functionality is centered around its consensus mechanism to allow a distributed network of nodes to agree on the common state of the ledger. Reaching agreement in such settings is even more difficult, particularly when combined with partial connectivity, dynamic membership, failures, and Byzantine behaviour. As blockchain networks continue to scale further, new solutions are needed that can preserve performance, security, and resilience at the extremes.

Over the years, various consensus protocols have been proposed for distributed systems and, more specifically, for blockchain, each tailored to different operating environments.
Classical consensus protocols such as Paxos, Raft, and Practical Byzantine Fault Tolerance (PBFT) have been shown to be effective in permissioned networks with trusted nodes and presumed reliable connectivity. However, these protocols typically rely on a stable leader or coordinator, leading to issues related to the presence of such a bottleneck and a single point of failure. 

Conversely, permissionless systems have moved towards probabilistic consensus algorithms such as Proof-of-Work (PoW) and Proof-of-Stake (PoS). Although these algorithms decentralise control and fault tolerance, they have critical limitations in terms of resource efficiency, latency, and vulnerability, as well as issues due to the tendency to centralisation of computational and/or financial power. 

Recent advances have proposed protocols based on epidemic-style sampling techniques, such as Avalanche and its variants. They try to enhance scalability and decentralisation by enabling nodes to achieve consensus via local interactions with randomly selected peers. 
Nevertheless, they typically incur costs of the trade-off between convergence time and communication overhead since frequent queries and parameter tuning are necessary in order to achieve a balance between responsiveness and efficiency. With increasing network size, these problems result in higher message complexity and slow convergence, 
restricting the use of such protocols in large and dynamic systems.

To address these issues, we present a novel consensus algorithm, Blockchain Epidemic Consensus Protocol (BECP), which is designed for large-scale decentralised environments. BECP is built on epidemic information dissemination and decentralised data aggregation to deliver a fully decentralised, leaderless consensus protocol. In BECP, nodes communicate with a randomly selected neighbour, achieving consensus through light-weight interactions. This keeps message complexity low, removes the need for a central leader or dense sampling, and possesses the property of fast convergence independent of message delays. 


Simulation results for the proposed protocol were evaluated on systems with up to 10,000 nodes, which validated ledger consistency and correctness. The evaluations were conducted under both normal network conditions and scenarios involving message delays. The results include measurements of the average number of calls to the fork resolution method, demonstrating the protocol’s effectiveness in handling block conflicts and its scalability in large networks. By combining the probabilistic guarantees of epidemic information spreading with efficient decentralised data aggregation techniques, the protocol achieves low-latency consensus with minimal message overhead. Extensive evaluations show that this approach significantly enhances communication efficiency compared to existing epidemic-based protocols, while maintaining comparable throughput and resilience to traditional consensus protocols.

The remainder of this paper is structured as follows. Section~\ref{sec:related_work} covers related work and current approaches to blockchain consensus. Section~\ref{sec:proposedMethod} details the design and the functionalities of BECP. Section~\ref{sec:implementation} outlines our implementation setup and the simulation environment. Section~\ref{sec:performance_evaluation} reports and analyses the performance results. Finally, Section~\ref{sec:conclusion} provides some conclusions and suggests future research directions.

\section{Related Work}
\label{sec:related_work}
Consensus protocols are the foundation of distributed networks, which allow nodes to achieve a shared state~\cite{hussein2023evolution}. In blockchain networks, they serve the essential function of consistency maintenance and conflict resolution. A good consensus protocol must provide for inclusive participation and efficient conflict resolution~\cite{singh2022survey}.

Consensus protocols can be classified into three general categories: direct communication-based, proof-of-x-based, and epidemic-based mechanisms~\cite{singh2022survey}. Direct communication-based protocols, i.e., Paxos~\cite{lamport2001paxos}, Raft~\cite{ongaro2014search}, and PBFT~\cite{lamport2019byzantine}, were initially designed for permissioned networks in which all nodes are authenticated. Paxos and Raft both follow leader-based models, though Paxos has more dynamic leadership changes and Raft has a more static leader election, whereas PBFT provides Byzantine fault tolerance by requiring a total of $3f + 1$ nodes to tolerate up to $f$ faulty nodes. This ensures that consensus can be achieved with at least $2f + 1$ consistent responses. These protocols have inspired many derivatives, such as MultiPaxos~\cite{chand2016formal}, Fast Paxos~\cite{lamport2006fast}, Byzantine Paxos~\cite{lamport2011byzantizing}, DBFT~\cite{neo2022whitepaper}, and FBA~\cite{singh2022survey}, which are being used extensively in blockchain ecosystems~\cite{lashkari2021comprehensive, ferdous2020blockchain, singh2022survey}.

Despite their original value, these classical protocols are best suited for permissioned and static contexts. Their adoption of deterministic periods, centralised leadership models, and complete connectivity assumptions did not prepare them for open, large-scale blockchain networks. Even recent enhancements like Kronos~\cite{liu2024kronos}, CollaChain~\cite{tennakoon2022collachain}, and MStableChain~\cite{li2024mstablechain} carry such architectural limitations with them, along with scalability and decentralisation problems.

As a response to centralisation, blockchain platforms have embraced proof-of-x-based consensus mechanisms. These decentralised protocols use cryptographic or economic proofs to validate blocks and settle forks—usually by choosing the longest chain (Nakamoto consensus)~\cite{nakamoto2008bitcoin} or the heaviest chain (GHOST)~\cite{sompolinsky2015secure}. The two most well-known representatives are Proof-of-Work (PoW)~\cite{nakamoto2008bitcoin, jakobsson1999proofs} and Proof-of-Stake (PoS)~\cite{king2012ppcoin, thin2018formal}. Although PoW provides strong security, it is limited by high energy consumption and hardware centralisation. 

PoS alleviates energy issues but brings with it risks like centralisation of stake and the "nothing-at-stake" issue, where participants are able to vote on several alternative versions of the blockchain simultaneously as they have nothing to lose—hence making consensus more difficult to achieve in a secure fashion.

A third, and emerging, category is epidemic-based consensus, which employs gossip-like spread of information and probabilistic agreement. The Avalanche and Snowman protocols in the Snow family of protocols demonstrate this approach by polling a random subset of nodes and deciding locally based on their votes~\cite{rocket2018snowflake, buchwald2024frosty}. These protocols excel in decentralisation and fault tolerance but are constrained by high message overhead due to repeated sampling and large sample sizes. An additional disadvantage is that nodes assume a full view of the network.

Alternative epidemic-based solutions concentrate on light-weight convergence. For instance, the Phase Transition Protocol (PTP)~\cite{ayiad2016agreement} and Epidemic Consensus Protocol (ECP)~\cite{ayiad2017agreement} employ random peer interaction and local computations to reach distributed agreement. These works demonstrate the power of epidemic models for decentralised networks.

Yet, current protocols in every category are limited. Classical protocols do not scale or decentralise. Proof-based solutions have issues with energy efficiency or economic equity. Existing epidemic protocols, while promising, tend to have high communication overhead. To alleviate these issues, we introduce the Blockchain Epidemic Consensus Protocol (BECP), a protocol based on epidemic communication principles and light local computations, without sampling. BECP provides adaptive and scalable consensus appropriate for blockchain networks, with resilience to message delays.

\section{The BECP Consensus Protocol}
\label{sec:proposedMethod}
The methodology followed here draws inspiration from the works in~\cite{ayiad2016agreement}, \cite{ayiad2017agreement}, \cite{kempe2003gossip}, and \cite{blasa2011symmetric} to develop a consensus protocol that is fully decentralised. The protocol is designed to run distributed computation on blockchain networks with no dependency on a known set of validators or leaders. In addition, it offers strong probabilistic assurances regarding convergence and optimises the efficiency of network resource utilisation.

BECP is a fully decentralised epidemic consensus protocol applied in blockchain technology contexts. BECP consists of three intertwined protocols that run in parallel: the System Size Estimation Protocol (SSEP)~\cite{ayiad2016agreement}, the Node Cache Protocol (NCP)~\cite{blasa2011symmetric}, and the Phase Transition Protocol (PTP)~\cite{ayiad2016agreement}. SSEP continuously monitors the system size, offering the function \textit{getSystemSize()}, which accurately estimates the number of participating nodes in a blockchain system in real time. Every node is initialised with a pair value $v$ and $w$, where $v=1$ and $w=1$ for a seed node and $w=0$ for any other node. 

NCP provides a scalable membership sampling function, \textit{getRandomNode()}, by randomly selecting nodes for epidemic communication purposes without complete knowledge of the system, which can be dynamic. NCP can only be used in simulations, as during the initialisation phase, each node randomly selects a set of neighbours from a global list of known node IDs, presuming unrealistic knowledge. In reality, the nodes can't know the neighbours in advance but can only directly connect to some other node. One possibility is to employ another membership protocol, such as EMP+, introduced in~\cite{poonpakdee2017robust}.

Finally, the PTP is a decentralised consensus algorithm that takes advantage of the other two protocols and eliminates duplicate issues and ensures the correct ordering of IDs within blocks. Similar to SSEP, in PTP two sets of estimators ($vp$,$wp$) and ($va$,$wa$) are maintained per block. The sets are utilised for estimating the number of nodes that received a proposal block utilising ($vp/wp$) in the Propagation phase and ($va/wa$) in the Agreement phase. The blocks go through the phases of propagation, followed by agreement, and then confirmation when the estimates are very close to the system size with a bounded error for a specified number of cycles~\cite{ayiad2016agreement}. 

The main benefit of comparing estimators to the system size is that it enables the detection of early convergence in the estimates $(vp/wp)$ and $(va/wa)$, improving throughput and consensus time~\cite{ayiad2016agreement}. It is an improvement over the Snow protocols' technique, which detects convergence after a longer time.

Like the NCP, nodes also have local caches (block local caches) that the PTP protocol uses during the consensus process. Nodes share their local caches and append new blocks obtained to their caches. Apart from these two local caches, nodes have a local ledger, which is a projection of the blockchain. They append confirmed blocks from the block local cache to their local ledgers. 

We make the following assumptions in our protocol: the physical network topology is connected and a trusted communication medium, i.e., messages are never lost, nodes are honest, and an active node remains on throughout the cycle by sending a Push message, getting a Push message, and answering with a Pull message.

The BECP protocol operates as follows. SSEP continuously produces an estimate, i.e., the system size or the total number of nodes participating, and this is subsequently compared with two estimations made by PTP for each block—specifically, the number of nodes that have received a block during each of its phases—in order to reach consensus. Simultaneously, NCP operates to network the nodes and balance information evenly among them, with the three protocols functioning in parallel during the same defined cycles, sharing the same Push and Pull messages.

Employment of the PTP protocol together with SSEP and NCP is not ideal for blockchain networks because of the fact that PTP reaches consensus on something that can or cannot be a chain of blocks (parents of blocks differ). This is because PTP does not take into consideration the parents or references of the blocks; it just resolves inconsistencies, duplicate blocks with the same ID. The nature and the structure of blockchain require that each item or block on which agreement is reached must refer to its parent (previous confirmed block). After creating a new block, its data and pointer cannot be changed since they are encrypted. This feature enables the protocol to generate a chain of interconnected blocks that is the same for all the participants.

The issue with the original PTP protocol is that blocks are being created and validated without corresponding references. One of the proposed solutions, though, is that participants create their blocks based on their previously created blocks; this introduces the risk of breaking the chain of references when a block is accepted from a group of candidate blocks having the same IDs. 

One other way of resolving this situation is to make nodes wait for the confirmation of the current latest block before creating their next blocks with reference to that specific block. Once a candidate block is selected as a confirmed block out of a group of other candidate blocks, the remaining candidates will be discarded from the local caches due to their invalid status. This method, while giving a sequence of blocks with accurate references, has high inefficiency as the nodes have to wait for the verification of blocks. Thus, this results in reducing the throughput of the systems.

In order to meet this challenge more effectively, we introduce a new consensus mechanism as outlined in Algorithm~\ref{alg:resolve_duplicate} to accept and resolve duplicate or inconsistent blocks, blocks with identical IDs. BECP's consensus mechanism has two distinct parts: duplicate blocks are resolved first in the local caches; afterwards, for the identified blocks, an estimation process is carried out. When the state of a block is changed to commit, which is done through the estimation process, the block is agreed upon. The second component, where the block is agreed upon, is slower. 

As shown in the algorithm, we introduce a new variable called $B_{pref}$, which represents the current preferred block. We impose nodes to create new blocks by referencing that block. Therefore, nodes have the ability to generate blocks irrespective of the verification of the previous block. The current preferred block $B_{pref}$ is the potential block that may be confirmed among a set of candidate blocks maintained in the local block cache of the nodes. A node, upon receiving a new preferred block, will include this new block while at the same time remove the current preferred block and its descendant blocks (Algorithm~\ref{alg:forkResolution}). Then, the node will generate a new block based on the newly preferred block as needed.

This method not only yields a steady sequence of blocks with accurate references but also enhances throughput because nodes do not await confirmation of the most recent block. Algorithm~\ref{alg:resolve_duplicate} is a modification of the Resolve Duplicate Block ID Procedure initially presented in PTP, utilised for selecting $B_{pref}$. Nodes initially carry out a comparison of the incoming block $\tau'$ with all the stored blocks in the block local cache $C_b$ for determining its duplication (line 2)—that is, if there is a stored block that has the same ID.

The duplication issue must be settled through the choice between the new block $\tau'$ and the current block $\tau$. Algorithm~\ref{alg:resolve_duplicate}, line 4, verifies whether $\tau'$ is identical to $\tau$. If they are identical, the process eliminates $\tau'$ while at the same time updating the pairs corresponding to $\tau$ based on its pairs $vp, wp, va,$ and $wa$. If $\tau'$ is not identical, the process then verifies whether $\tau'$ had been labeled as $B_p$, as indicated in line 6. If so, the process performs a fork resolution process, removes $\tau$, inserts $\tau'$ into block local cache $C_b$, and sets $B_{pref}$ to $\tau'$. $\tau'$ is selected as the temporary block $B_{pref}$ under either case ($\tau'.t=\tau.t$ and $\tau'.o<\tau.o$) or under case ($\tau'.t<\tau.t$) provided $\tau$ is yet to be confirmed, meaning selecting a temporary block based on block generation time ($t$) and proposer ID ($o$).
\begin{algorithm}[h!]
\caption{Revised Resolve Duplicate Block ID Procedure}
\label{alg:resolve_duplicate}
\footnotesize
\begin{algorithmic}[1]
\raggedright
\ForAll{$\tau' \in m.C_b$}
    \If{$C_b$ contains $\tau$ such that $\tau'.id = \tau.id$}
        \If{$\tau$ has not been confirmed}
            \If{$\tau'.t = \tau.t$ \textbf{and} $\tau'.o = \tau.o$}
                \State $\tau \gets \langle \tau.id, \tau.o, \tau.t, \tau.vp + \tau'.vp, \tau.wp + \tau'.wp, \tau.va + \tau'.va, \tau.wa + \tau'.wa, \tau.\text{state} \rangle$
            \ElsIf{$\tau'.t = \tau.t$ \textbf{and} $\tau'.o < \tau.o$ \textbf{or} $\tau'.t < \tau.t$}
                \State \textsc{forkResolution}($C_b$, $\tau$)
                \State $\tau \gets \langle \tau'.id, \tau'.o, \tau'.t, \tau'.vp + 1, \tau'.wp, \tau'.va, \tau'.wa, \tau'.\text{state} \rangle$
                \State Set $B_{pref} \gets \tau'$
            \EndIf
        \EndIf
    \ElsIf{creator of parent of $\tau'$ = creator of $B_{pref}$}
        \State $C_b \gets C_b \cup \{\langle \tau'.id, \tau'.o, \tau'.t, \tau'.vp + 1, \tau'.wp, \tau'.va, \tau'.wa, \tau'.\text{state} \rangle\}$
        \State Set $B_{pref} \gets \tau'$
    \EndIf
\EndFor
\end{algorithmic}
\end{algorithm}
\begin{algorithm}[h!]
\caption{Fork Resolution Procedure}
\label{alg:forkResolution}
\footnotesize
\begin{algorithmic}[1]
\Statex \textbf{Data:} Received Block $\tau'$
\Statex \textbf{Result:} Discarding Invalid Blocks
\Function{\textsc{forkResolution}}{$C_b$, $\tau'$}
    \State $Ch \gets \tau'.\text{getChildren()}$
    \ForAll{$ch \in Ch$}
        \State \textit{\textsc{forkResolution}($C_b$, $ch$)}
    \EndFor
    \State remove $\tau'$ from the $C_b$
\EndFunction
\end{algorithmic}
\end{algorithm} 
\begin{algorithm}[h!]
\caption{Block Generation Procedure}
\label{alg:blockgen}
\footnotesize
\begin{algorithmic}[1]
\Statex \textbf{Result:} Generate a new block $B_b$
\Function{generateNewBlock}{ }
    \If{$\forall B \in C_b, \; B.id \neq B_{p}.id + 1$} 
        \State generate a new block $B_b$
        \State $Ch_{B_p} \gets Ch_{B_p} \cup \{B_b\}$
        \State $C_b \gets C_b \cup \{B_b\}$
        \State \text{set} $B_p$ to $B_b$
    \EndIf \Comment{Otherwise, wait until the block is either dropped or selected as a $B_p$.}
\EndFunction
\end{algorithmic}
\end{algorithm}

\begin{figure}[!t]
\centering
\subfloat[Case I]{\includegraphics[clip, trim=2mm 2mm 2mm 2mm
, width=3.5in]{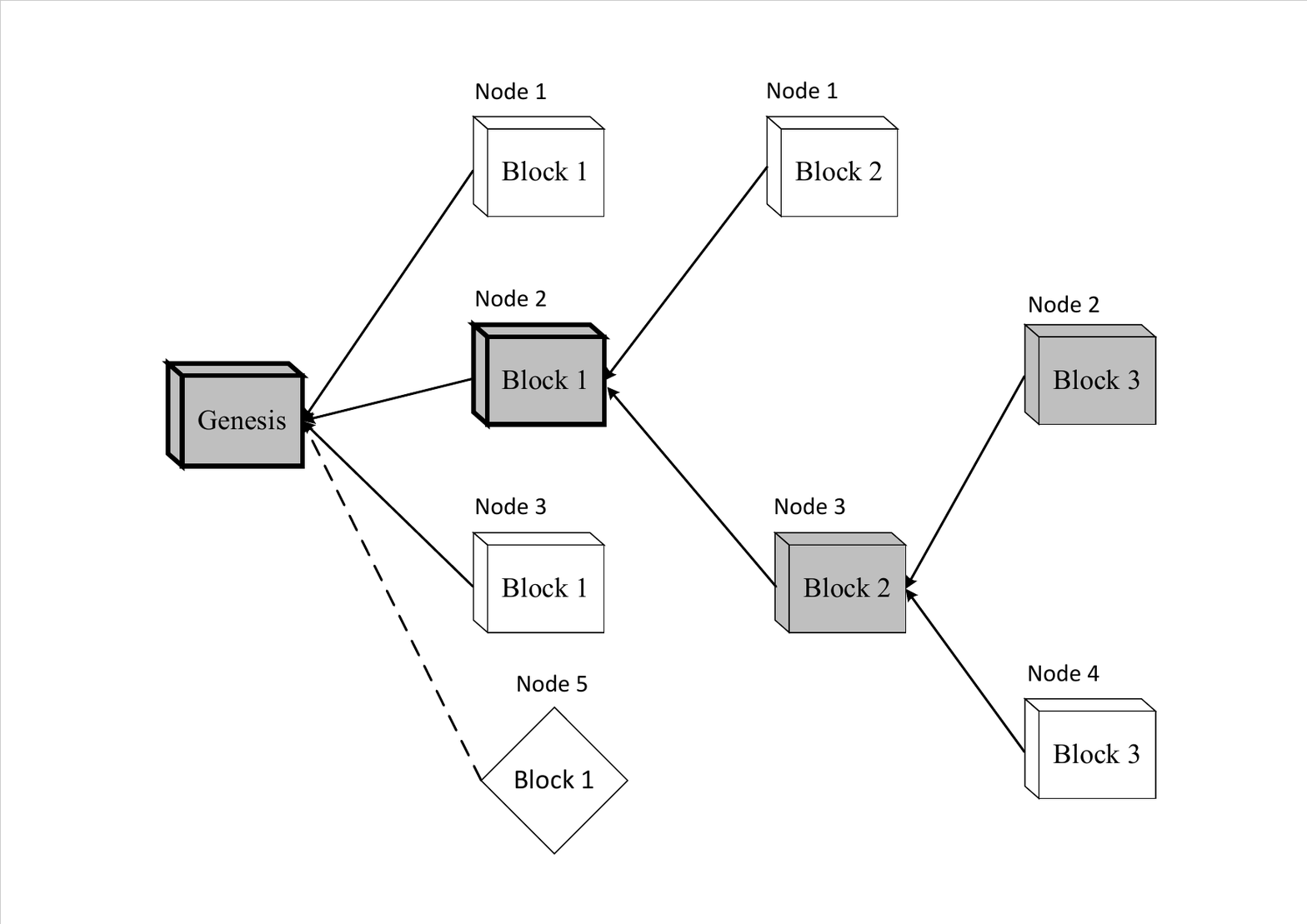}
\label{fig:immutability}}
\hspace{1in}
\subfloat[Case II]{\includegraphics[clip, trim=2mm 2mm 2mm 2mm
, width=3.5in]{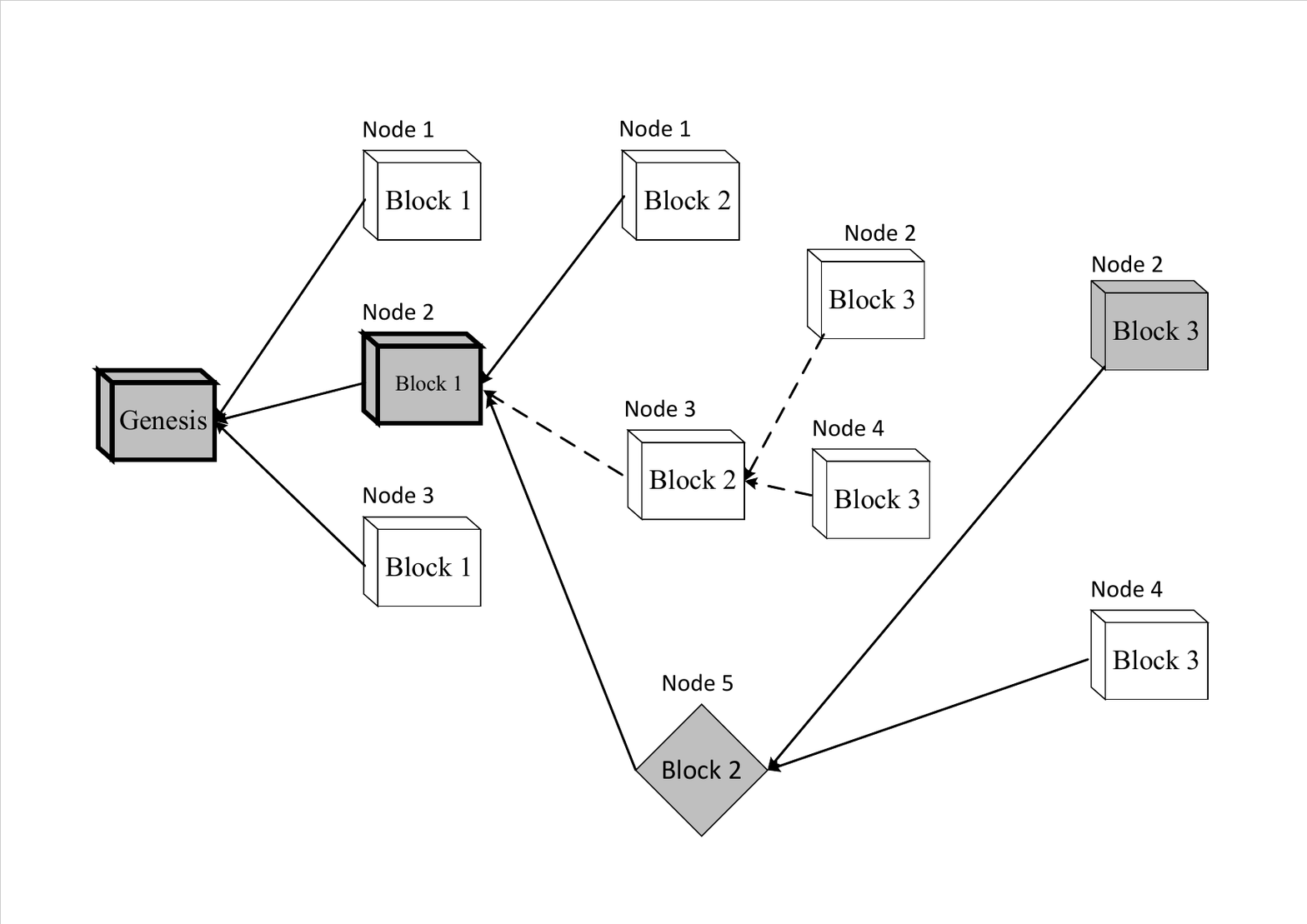}
\label{fig:forkResolution}}
\caption{Illustration of Protocol Immutability (Case I), Fork Resolution (Case II): White blocks represent candidate blocks that dropped, pale blocks denote the current preferred blocks, the diamond block indicates the newly received block identified as the new preferred block, and blocks with highlighted edges signify confirmed blocks. The text over the blocks indicates the node that created each block.}
\end{figure}

Figures~\ref{fig:immutability} and \ref{fig:forkResolution} represent two different cases which illustrate distinct types of outcomes involving block resolution and handling in the system. For Case I, the system runs normally. However, in this example, the diamond-shaped block marked by the ID of 1 is illustrated as having arrived after other blocks. Since block 1 has already been confirmed—the block with the highlighted edges—nodes go on to ignore the newly arrived diamond-shaped block.

Case II illustrates a fork resolution case. Block 2, generated by node 3, has been identified as a preferred block. Upon receiving the diamond-shaped block 2, nodes call upon a process of fork resolution, during which the already chosen preferred block 2, generated by node 3, and its descendants are invalidated and discarded, leading to the consideration of the diamond-shaped block as a new preferred block. Subsequent blocks are then referred towards this newly designated preferred block.

Moreover, we introduce a new code snippet (Algorithm~\ref{alg:blockgen}) dedicated to the block generation process, which prevents nodes from double-spending and creating blocks with identical IDs and creators. In the algorithm, nodes verify whether there are no already existing blocks with identical IDs in their local caches, and the new block references the current preferred block $B_{pref}$. 

In detail, nodes add new blocks into the local cache $C_b$ as follows: They first verify that there is no other block of a greater ID than $B_{pref}$. Correspondingly, if so, they generate a new block, refer it to the current $B_{pref}$—the block with the highest ID—insert the new block in the children set of $Ch_{B_{pref}}$ and into the block local cache $C_b$, and they finally set the new block as a new $B_{pref}$. Otherwise, they wait until either the block is dropped or is chosen as a new $B_{pref}$. This is a precautionary measure to keep the blocks in the proper order.

\section{Implementation and Network Configuration}
\label{sec:implementation}
In order to make a comparison between the performance of the BECP protocol and other studied blockchain consensus protocols, a suitable simulation platform must be adopted. For this purpose, the Just Another Blockchain Simulator (JABS)~\cite{yajam2023jabs} was selected due to its modularity and useful characteristics. In the following, we introduce the simulator.

\subsection{JABS Blockchain Simulator}
JABS is a free, open-source, discrete-event Java simulator specifically designed for simulating large-scale blockchain networks with up to 10,000 nodes in close-to-realistic environments. It simulates different consensus mechanisms and network topologies and hence is a great platform for comprehensive performance and security analysis of blockchain protocols.

In JABS, an event occurs when a node receives a message from any other peer. Simulation starts by running initial events in which each node creates a block, timestamps it, and passes it to its neighbour. Nodes keep producing new blocks at fixed intervals according to a predefined probability. Block and transaction sizes are taken from empirical distributions using real Bitcoin data. Simulation halts when there are no longer any events being generated or when a specified simulation time has elapsed.

There is no pre-defined loop or cycle of execution in the current implementation of JABS. Thus, a new event type was introduced to support periodic node activation—a prerequisite for BECP, Avalanche, Snowman, and Raft. It allows nodes to prepare and plan the next activation cycle in each simulation cycle.

\subsection{Protocol and Network Configuration Parameters}
The simulator was utilised to implement and experiment with a collection of consensus protocols: PAXOS, RAFT, PBFT, BECP, Avalanche, and Snowman. Simulations were run over network sizes from 1,000 to 10,000 nodes for a duration of 300 seconds: the extremely high complexity of Avalanche, Snowman, and PBFT capped their scale at 5,000 nodes. For scalability testing, BECP was particularly tested up to 10,000 nodes. All simulations were run with different random seeds on high-performance computing facilities.

All protocols were evaluated in a Wide Area Network (WAN) topology. Message latency was simulated as either uniformly distributed (with a minimum of 0.05 seconds and a maximum of 0.15 seconds) or Pareto distributed with constant scale parameter $x_m = 0.05$ and shape parameter $\alpha$ ranging from 4 to 8, to allow simulation of heavy-tailed latency distributions typical in decentralised networks. The priority in this work is the analysis of consensus protocols; thus, the details of transaction generation were abstracted. We assumed that transactions are gathered by proposer nodes and blocks are created accordingly.

In deterministic protocols like PAXOS, RAFT, and PBFT, block creation is leader-driven—blocks are proposed only after consensus on the prior block is reached. However, in probabilistic protocols like BECP, Avalanche, and Snowman, nodes propose blocks in intervals without waiting for consensus on prior blocks. This approach more accurately captures real-world blockchain system dynamics. For BECP, Avalanche, and Snowman, the block generation time ($T_{block}$) was set to 10 seconds, and the block generation probability ($P_{block}$) was 5\%. The same $T_{block}$ was used for other protocols for consistency. Node activation mechanisms differed across protocols. BECP, Avalanche, Snowman, and Raft employed periodic activation cycles (0.351 seconds), whereas PAXOS, RAFT, and PBFT used message-driven activation initiated by the leader.

Protocol-specific parameters were also set. For BECP, the thresholds for estimation error ($\epsilon$), minimum consecutive cycles ($\Psi$), and neighbour cache size ($N_{cache}$) are the parameters. Avalanche and Snowman parameters are: sample size ($k$), quorum threshold ($\alpha_1$), early commitment threshold ($\beta_1$), and consecutive threshold ($\beta_2$). $\alpha_2$ is the finalisation threshold of a block in Snowman. Avalanche was instantiated with a chain structure (instead of its native DAG) to allow a fair comparison with the other blockchain protocols. In this setup, new blocks pointed to the highest-ID block in the known set, allowing for a linear chain topology. Table~\ref{table:experiment_settings} provides a comprehensive overview of all simulation parameters employed.
\begin{table*}[h!]
  \setlength{\tabcolsep}{5pt} 
  \renewcommand{\arraystretch}{1.1} 
  \caption{Experiment Settings}
  \label{table:experiment_settings}
  \centering
  \footnotesize
  \begin{tabular}{c c c c c c c}
    \hline
    \textbf{Parameters} & \textbf{BECP} & \textbf{Avalanche} & \textbf{PBFT} & \textbf{PAXOS} & \textbf{RAFT} & \textbf{Snowman} \\
    \hline
    $D_1$ (s) & 0.05 & 0.05 & $-$ & $-$ & $-$ & 0.05 \\
    Network Latency (s) & [0.05, 0.15) & [0.05, 0.15) & [0.05, 0.15) & [0.05, 0.15) & [0.05, 0.15) & [0.05, 0.15)\\
    Cycle Time (s) & 0.35 & 0.35 & $-$ & $-$ & 0.35 & 0.35\\
    $K$ & $-$ & 20 & $-$ & $-$ & $-$ & 20\\
    $\alpha_1$ & $-$ & 0.8 & $-$ & $-$ & $-$ & 10\\
    $\alpha_2$ & $-$ & $-$ & $-$ & $-$ & $-$ & 15\\
    $\beta_1$ & $-$ & 50 & $-$ & $-$ & $-$ & 15 \\
    $\beta_2$ & $-$ & 150 & $-$ & $-$ & $-$ & $-$ \\
    $T_{\text{block}}$ (s) & 10 & 10 & 10 & 10 & 10 & 10\\
    $P_{\text{block}}$ & 5\% & 5\% &  $-$ &  $-$ &  $-$ & 5\%\\
    $\epsilon$ & 0.05 & $-$ & $-$ & $-$ & $-$ & $-$\\
    $N_{\text{cache}}$ & 100 & $-$ & $-$ & $-$ & $-$ & $-$ \\
    $\Psi$ & 5 & $-$ & $-$ & $-$ & $-$ & $-$\\
    Timeout Range & $\psi = 1, \tau = 2$ & $-$ & $-$ & $-$ & $1.0$ to $1.2$ & $-$\\
    \hline
  \end{tabular}
\end{table*}

\subsection{Correctness and Delay Scenario}
In order to test protocol robustness under delay conditions, we used a Pareto distribution with parameters ($x_m = 0.05$, $\alpha = 4$ to $8$) to model heavy-tailed message delays in these experiments. Lower $\alpha$ values add more delay variability. The simulations were executed for a duration of 300 seconds. A test function to verify correctness was implemented to ensure blockchain connectivity among all nodes by following block hashes. An experiment is considered to be a pass only if it successfully runs through all five trials. A test fails if any one of the trials fails. 

\section{Performance Evaluation}
\label{sec:performance_evaluation}
In this section, the performance metrics utilised to compare consensus protocols are explained first. We then present the performance of PBFT, RAFT, Paxos, BECP, Snowman, and Avalanche under normal network conditions, i.e., without introducing simulated delays for messages. Afterwards, we present BECP, Snowman, and Avalanche simulation results under a scenario where message delays are introduced.

\subsection{Evaluation Measures}
Consensus protocols are generally characterised in terms of a number of crucial performance metrics such as throughput, scalability, latency, and communication overhead. \textbf{Throughput} measures the number of proposals that are agreed upon per second. Higher throughput generally indicates an efficient protocol, particularly useful in cryptocurrency networks where the speed of transaction processing affects usability and adoption. 
\textbf{Scalability} is a measure of how effectively the protocol will function when the network is large. While real-world systems have the prospect of participants growing over time, scalability is required to provide long-term sustainability.
\textbf{Latency} is the time difference between the creation of a proposal and its verification by consensus. Reducing this delay is important for real-world applications where users anticipate quick confirmation of their action. 
\textbf{Communication Overhead} is the total number of messages sent between nodes during the simulation period. Low overhead is preferable, since too much messaging can cause network congestion and decreased overall efficiency.

Table~\ref{tab:fork_calls_varying_nodes} and Table~\ref{tab:fork_calls_varying_prob} show the average number of \textit{forkResolution} calls per block per node in BECP. According to the results, as the network size increases, the average number of fork resolution calls also increases, although not always in a strictly linear manner. This indicates that although larger systems have more forks, the fork resolution mechanism does scale correctly. Furthermore, as the block generation probability rises, the fork resolution calls rise dramatically—implying a greater chance of concurrent block generation. Nonetheless, the rate of throughput and average consensus time remain fixed at high probabilities, implying that the protocol handles forks effectively even at high generation rates. These results confirm BECP's capacity to manage forks without compromising scalability and performance.
\begin{table}[h!]
    \centering
    \small
    \caption{Average calls to fork resolution method per block in each node for varying number of nodes}
    \label{tab:fork_calls_varying_nodes}
    \begin{tabular}{cc}
        \toprule
        \multicolumn{2}{c}{\textbf{Average Calls to Fork Resolution Method per Block in Each Node}} \\
        \midrule
        \multicolumn{2}{l}{\textbf{Simulation time:} 300 seconds} \\
        \multicolumn{2}{l}{\textbf{Block Generation Probability:} 0.05} \\
        \midrule
        \textbf{Number of Nodes} & \textbf{BECP} \\
        \midrule
        1000  & 2.722 \\
        2000  & 3.1472 \\
        4000  & 3.6294 \\
        6000  & 3.8106 \\
        8000  & 4.0015 \\
        10000 & 4.1096 \\
        \bottomrule
    \end{tabular}
\end{table}
\begin{table}[h!]
    \centering
    \small
    \caption{BECP performance under varying block generation probabilities. Values show: avg. fork resolution calls per block per node / blocks reaching consensus / avg. consensus time (seconds).}
    \label{tab:fork_calls_varying_prob}
    \begin{tabular}{cc}
        \toprule
        \textbf{Block Generation Probability} & \textbf{BECP Metrics (calls / blocks / time)} \\
        \midrule
        0.05  & 4.1096 / 29 / 10.4800 \\
        0.2   & 5.0653 / 29 / 10.4745 \\
        0.4   & 5.5992 / 29 / 10.4847 \\
        0.6   & 5.9332 / 29 / 10.4866 \\
        0.8   & 6.1591 / 29 / 10.4760 \\
        1     & 6.3507 / 29 / 10.4599 \\
        \bottomrule
    \end{tabular}
\end{table}
\subsection{Performance Under Normal Network Conditions}
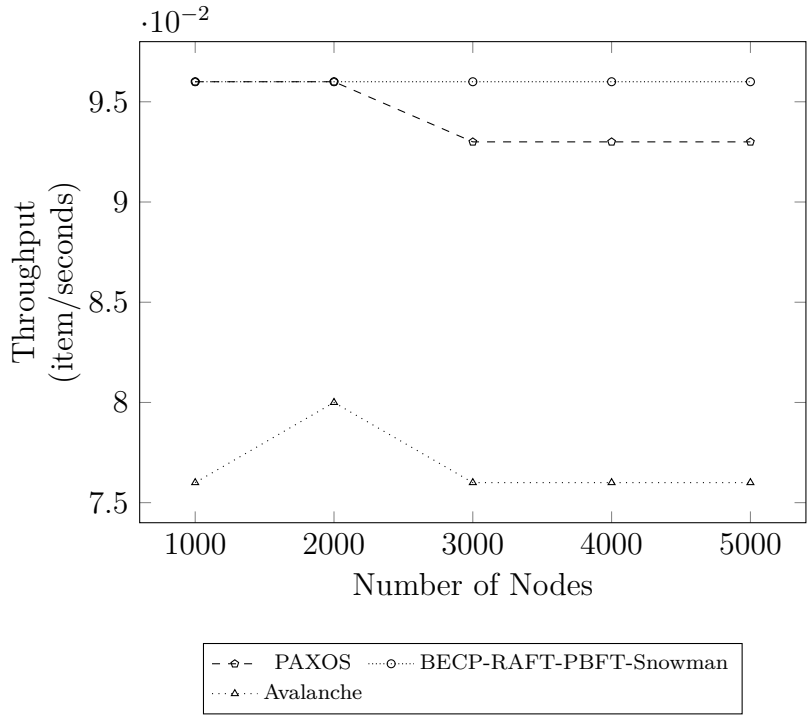
\begin{figure}
    \centering
    \begin{tikzpicture}
        \begin{axis}[
            width=0.65\textwidth,            height=0.5\textwidth,
            xlabel={Number of Nodes},
            ylabel={\shortstack{Throughput\\ (item/seconds)}},
            scaled ticks=true,
            tick label style={/pgf/number format/fixed},
            legend style={at={(0.5,-0.25)},
            anchor=north,legend columns=2, font=\scriptsize},
            symbolic x coords={1000, 2000, 3000, 4000, 5000},
            xtick=data,
            mark size=1.5,
            mark options={solid},
        ]
        \addplot[dashed, mark=pentagon] coordinates {(1000, 0.096) (2000, 0.096) (3000, 0.093) (4000, 0.093) (5000, 0.093)};
        \addplot[densely dotted, mark=o] coordinates {(1000, 0.096) (2000, 0.096) (3000, 0.096) (4000, 0.096) (5000, 0.096)};
        \addplot[dotted, mark=triangle] coordinates {(1000, 0.076) (2000, 0.08) (3000, 0.076) (4000, 0.076) (5000, 0.076)};
        \legend{PAXOS, BECP-RAFT-PBFT-Snowman, Avalanche}
        \end{axis}
    \end{tikzpicture}
    \caption{Comparative Throughput Analysis: BECP, RAFT, PAXOS, PBFT, Snowman, and Avalanche for node counts from 1000 to 5000. RAFT, PBFT, Snowman, and BECP have identical throughput across all node counts.}
    \label{fig:throughput}
\end{figure}

Figures~\ref{fig:throughput}, \ref{fig:overhead}, and \ref{fig:latency} compare PAXOS, RAFT, PBFT, Snowman, Avalanche, and BECP performance results. As evident from Figure~\ref{fig:throughput}, RAFT, PBFT, Snowman, and BECP have similar throughput, whereas PAXOS is somewhat lower because it has an additional promise phase in every round, which introduces latency. RAFT optimises this by electing a leader just once. BECP has slightly better throughput than Avalanche because of earlier consensus detection achieved by comparing the system size with the number of nodes that have received a block in the system, and also through epidemic communication that utilises both push and pull mechanisms.

Figure~\ref{fig:overhead} indicates that PBFT incurs the highest communication overhead due to its all-to-all voting for each round. RAFT also incurs maximum overhead due to the extra heartbeat messages from its leader. PAXOS, being more centralised, incurs less overhead as only the leader broadcasts blocks. BECP, Snowman, and Avalanche, being decentralised, incur more overhead than PAXOS as nodes can propose block creation without waiting for confirmations. Among these, BECP has less overhead than Snowman and Avalanche since each BECP node communicates to one peer instead of sampling from $K$ peers, which exponentially grows the number of messages in the others.

In terms of latency, the results in Figure~\ref{fig:latency} demonstrate that the classical protocols' determinism leads to low latency, finishing consensus in a couple of message exchanges, apart from leader election. However, they handle blocks sequentially. BECP, Snowman, and Avalanche, allowing probabilistic block creation, have higher latency but more accurately reflect real-world asynchronous environments. BECP outperforms Snowman and Avalanche on latency while matching the classical protocols' performance on throughput. Simulations with network sizes ranging from 1,000 to 5,000 nodes were performed, which show that BECP scales favourably with higher throughput and lower overhead compared to Avalanche, PBFT, RAFT, and Snowman.

Figures~\ref{fig:BECPthroughput}, \ref{fig:BECPoverhead}, and \ref{fig:BECPlatency} demonstrate BECP's scalability to networks of sizes of up to 10,000 nodes. Throughput is constant at around 0.096 blocks per second, irrespective of network size. The overall communication overhead is still logarithmic in size. For latency, BECP still maintains an average of 10 seconds for all network sizes, confirming its effectiveness and feasibility in large-scale settings. Simulations of all the protocols were tested with the test function, which confirmed their correct operation.
\begin{figure}[h!]
    \centering
    \begin{tikzpicture}
        \begin{axis}[
            width=0.65\textwidth,
            height=0.5\textwidth,
            xlabel={Number of Nodes},
            ylabel={\shortstack{Number of Sent\\Messages}},
            ymode=log,
            log basis y={10},
            legend style={at={(0.5,-0.25)},
            anchor=north,legend columns=4, font=\scriptsize},
            symbolic x coords={1000, 2000, 3000, 4000, 5000},
            xtick=data,
            mark size=1.5, 
            mark options={solid}, 
        ]
        \addplot[dashed, mark=pentagon] coordinates {(1000, 144650) (2000, 288959) (3000, 432954) (4000, 576993) (5000, 723839)};
        \addplot[solid, mark=x] coordinates {(1000, 1709670) (2000, 3419364) (3000, 5129016) (4000, 6838714) (5000, 8548385)};
        \addplot[dashed, mark=star] coordinates {(1000, 1939000) (2000, 5878000) (3000, 11817000) (4000, 19756000) (5000, 29695000)};
        \addplot[dashdotted, mark=square] coordinates {(1000, 59092194) (2000, 236490591) (3000, 531941979) (4000, 947090995) (5000, 1478811660)};
        \addplot[dotted, mark=triangle] coordinates {(1000, 34193533) (2000, 68387173) (3000, 102580719) (4000, 136774297) (5000, 170967792)};
        \legend{PAXOS, BECP, RAFT, PBFT, Snowman-Avalanche}
        \end{axis}
    \end{tikzpicture}
    \caption{Comparative communication overhead of BECP, RAFT, PAXOS, PBFT, Snowman, and Avalanche for 1,000–5,000 nodes. A single line represents Snowman and Avalanche due to identical overhead across node counts.}
    \label{fig:overhead}
\end{figure}
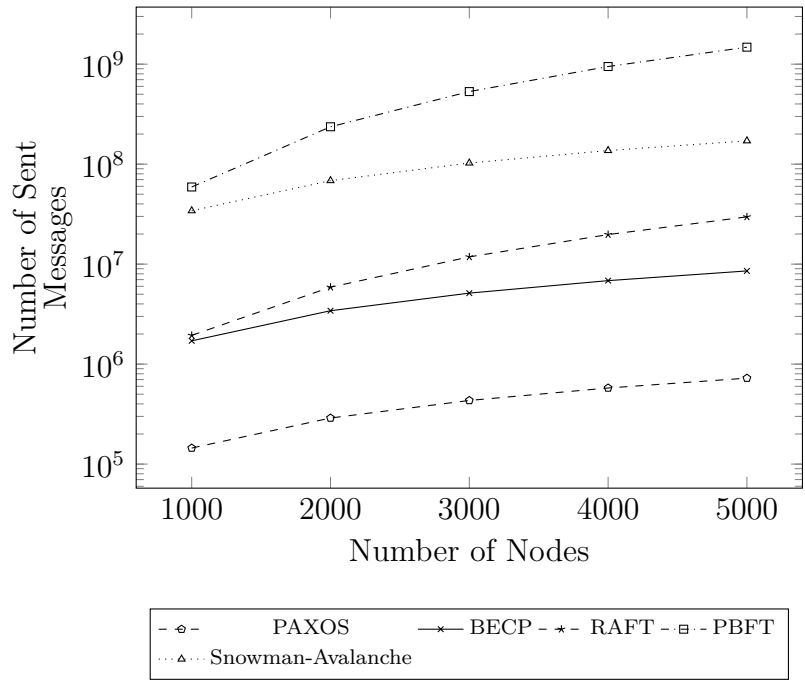
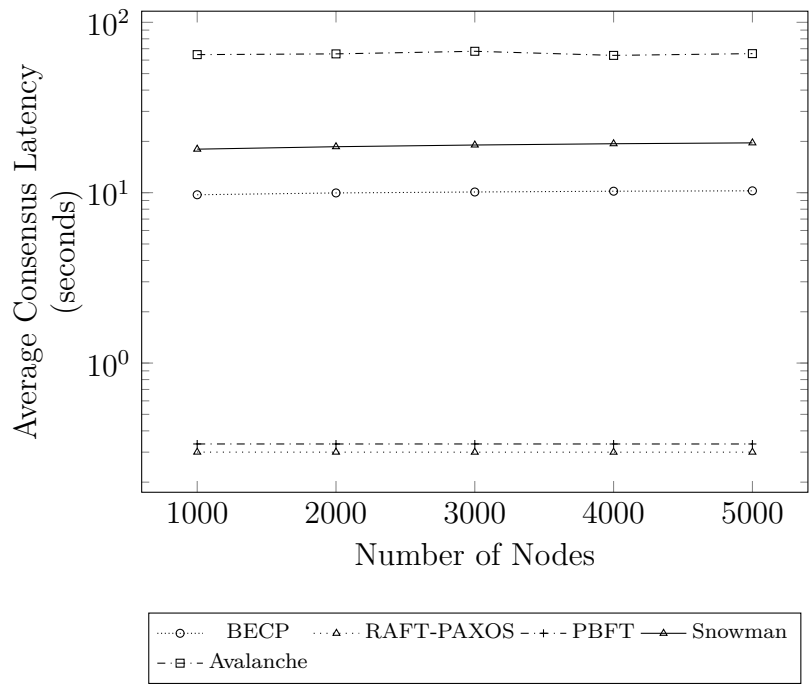
\begin{figure}[h!]
    \centering
    \begin{tikzpicture}
        \begin{axis}[
            width=0.65\textwidth,
            height=0.5\textwidth,
            xlabel={Number of Nodes},
            ylabel={\shortstack{Average Consensus Latency\\(seconds)}},
            ymode=log,
            log basis y={10},
            legend style={at={(0.5,-0.25)},
            anchor=north,legend columns=4, font=\scriptsize},
            symbolic x coords={1000, 2000, 3000, 4000, 5000},
            xtick=data,
            mark size=1.5, 
            mark options={solid}, 
        ]
        \addplot[densely dotted, mark=o] coordinates {(1000, 9.7234) (2000, 9.9666) (3000, 10.0936) (4000, 10.2019) (5000, 10.2545)};
        \addplot[dotted, mark=triangle] coordinates {(1000, 0.3001) (2000, 0.3) (3000, 0.2999) (4000, 0.2999) (5000, 0.3)};
        \addplot[dashdotted, mark=+] coordinates {(1000, 0.3351) (2000, 0.335) (3000, 0.3351) (4000, 0.335) (5000, 0.335)};
        \addplot[solid, mark=triangle] coordinates {(1000, 18.0078) (2000, 18.6170) (3000, 19.0484) (4000, 19.3792) (5000, 19.6071)};
        \addplot[dashdotted, mark=square] coordinates {(1000, 64.6251) (2000, 65.2678) (3000, 67.6082) (4000, 63.9532) (5000, 65.5027)};
        \legend{BECP, RAFT-PAXOS, PBFT, Snowman, Avalanche}
        \end{axis}
    \end{tikzpicture}
    \caption{Comparative analysis of average consensus latency for node counts ranging from 1,000 to 5,000. PAXOS and RAFT are represented by a single line due to identical consensus latency across all node counts.}
    \label{fig:latency}
\end{figure}
\subsection{BECP Performance under Message Delays}
Table~\ref{table:delayTests} compares BECP, Snowman, and Avalanche under simulated message delays, in terms of throughput and test function outcomes. BECP has stable performance across different $\alpha$ values, which means that it is resilient to delays in push/pull messaging. Even when $\alpha$ is lowered from 8 to 4, there is no throughput loss for BECP. While Snowman also performs well, it requires a lot more messages than BECP. Table~\ref{table:BECP_SSEP_Delay} represents BECP's performance for a network of 10,000 nodes under delay. BECP passed the test function at all $\alpha$ values that we tested, confirming stable performance.
\begin{figure}[h!]
    \centering
    \begin{tikzpicture}
        \begin{axis}[
            width=0.65\textwidth,
            height=0.5\textwidth,
            xlabel={Number of Nodes},
            ylabel={\shortstack{Throughput\\ (item/seconds)}},
            scaled ticks=true,
            tick label style={/pgf/number format/fixed},
            legend style={at={(0.95,0.05)},
            anchor=south east,legend columns=1, font=\scriptsize},
            symbolic x coords={1000, 5000, 10000},
            xtick=data,
            mark size=1.5,
            mark options={solid},
        ]
        \addplot[solid, mark=pentagon] coordinates {(1000, 0.096) (5000, 0.096) (10000, 0.096)};
        \legend{BECP}
        \end{axis}
    \end{tikzpicture}
    \caption{Analysis of throughput, measured in blocks per second, for node counts ranging from 1,000 to 10,000 in BECP}
    \label{fig:BECPthroughput}
\end{figure}
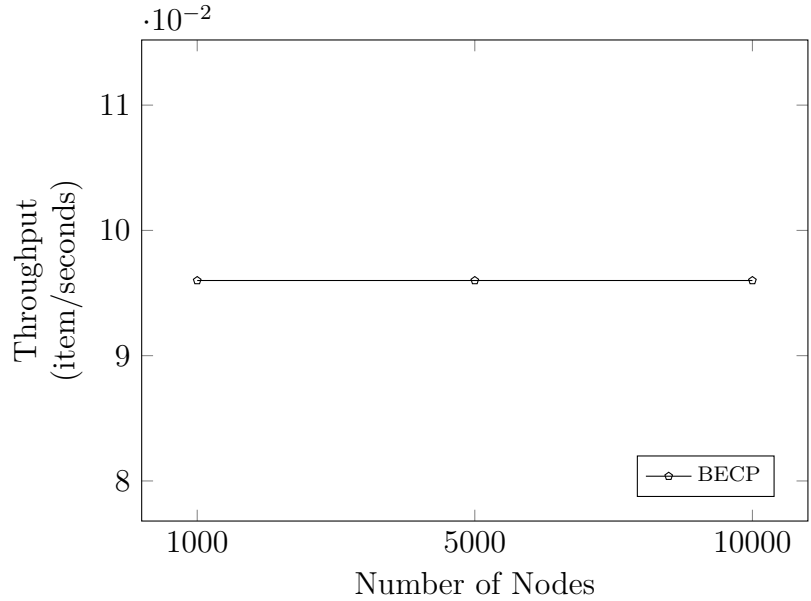
\begin{figure}[h!]
    \centering
    \begin{tikzpicture}
        \begin{axis}[
            width=0.65\textwidth,
            height=0.5\textwidth,
            xlabel={Number of Nodes},
            ylabel={\shortstack{Number of Sent\\Messages}},
            ymode=log,
            log basis y={10},
            legend style={at={(0.95,0.05)},
            anchor=south east,legend columns=1, font=\scriptsize},
            symbolic x coords={1000, 5000, 10000},
            xtick=data,
            mark size=1.5, 
            mark options={solid}, 
        ]    
        \addplot[solid, mark=pentagon] coordinates {(1000, 1709670) (5000, 8548385) (10000, 17096769)};
        \legend{BECP}
        \end{axis}
    \end{tikzpicture}
    \caption{Analysis of communication overhead, measured in total number of sent messages during the simulation period, for node counts ranging from 1,000 to 10,000 in BECP}
    \label{fig:BECPoverhead}
\end{figure}
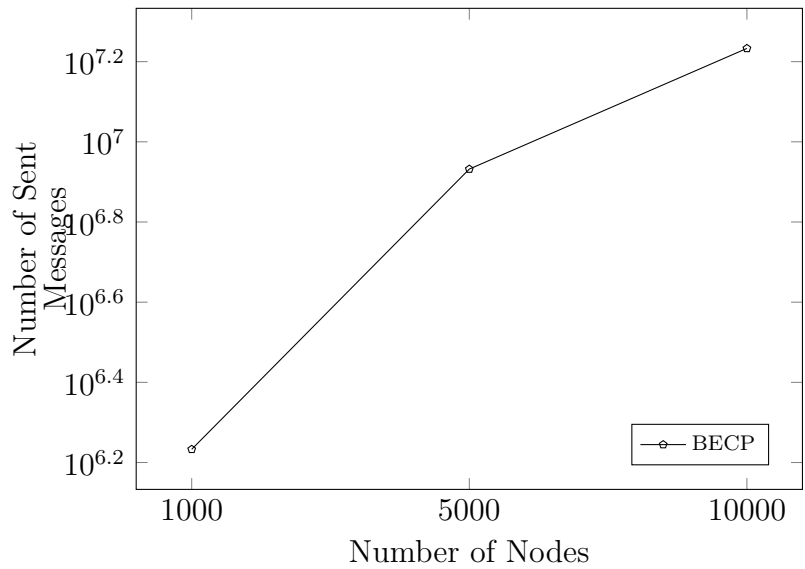
\begin{figure}[h!]
    \centering
    \begin{tikzpicture}
        \begin{axis}[
            width=0.65\textwidth,
            height=0.5\textwidth,
            xlabel={Number of Nodes},
            ylabel={\shortstack{Average Consensus Latency\\(seconds)}},
            legend style={at={(0.95,0.05)},
            anchor=south east,legend columns=1, font=\scriptsize},
            symbolic x coords={1000, 5000, 10000},
            xtick=data,
            mark size=1.5, 
            mark options={solid}, 
            ymin=0,  
            ymax=20, 
        ]
        \addplot[solid, mark=pentagon] coordinates {(1000, 9.7234) (5000, 10.2545) (10000, 10.4800)};
        \legend{BECP}
        \end{axis}
    \end{tikzpicture}
    \caption{Analysis of average consensus latency for node counts ranging from 1,000 to 10,000 in BECP}
    \label{fig:BECPlatency}
\end{figure}
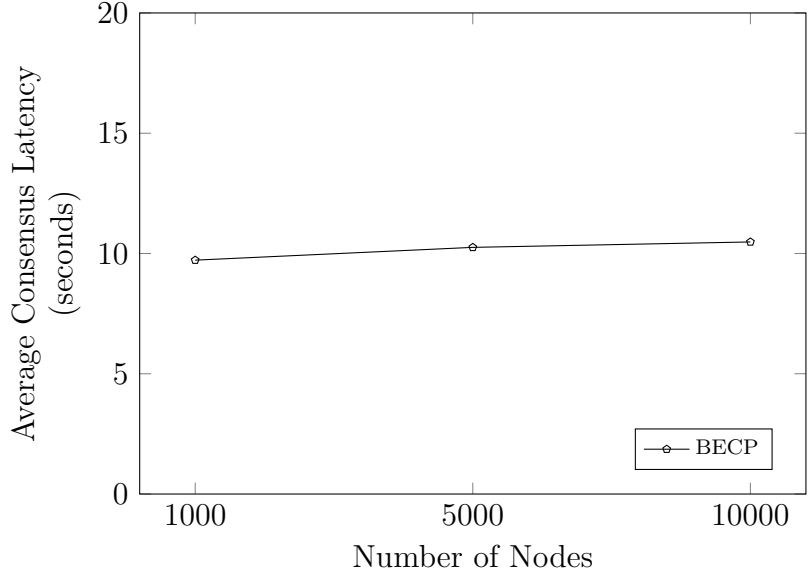
\begin{table}[h!]
\centering
\begin{tabular}{p{0.5cm} p{2.5cm} p{2.5cm} p{2.5cm}}
\hline
\textbf{$\alpha$} & \textbf{BECP} & \textbf{Snowman} & \textbf{Avalanche} \\ \hline
8 &Failed(0)/29&Failed(0)/29 &Failed(1)/24\\ 
7 &Failed(0)/29&Failed(0)/29&Failed(0)/24\\ 
6 &Failed(0)/29&Failed(0)/29&Failed(0)/24\\ 
5 &Failed(0)/29&Failed(0)/29&Failed(1)/23 \\ 
4 &Failed(0)/29&Failed(0)/29&Failed(1)/23\\ \hline
\end{tabular}
\caption{Results of a simulated message delay scenario with 5,000 nodes in a 5-minute simulation}
\label{table:delayTests}
\end{table}
\begin{table}[H]
\small
\centering
\begin{tabular}{@{}c c c c c c@{}}
\hline
\textbf{$\alpha$} & 8 & 7 & 6 & 5 & 4 \\ \hline
\textbf{BECP} & Failed(0)/29 & Failed(0)/29 & Failed(0)/29 & Failed(0)/29 & Failed(0)/29 \\ \hline
\end{tabular}
\caption{Results of a simulated message delay scenario for BECP with 5,000 nodes in a 5-minute simulation}
\label{table:BECP_SSEP_Delay}
\end{table}
\section{Conclusion}
\label{sec:conclusion}
This paper presents a novel scalable consensus protocol, Blockchain Epidemic Consensus Protocol (BECP). BECP is a fully decentralised protocol designed for large-scale blockchain systems, which adopts epidemic communication and local processing to reach consensus on blocks. In contrast to classical protocols like PAXOS, RAFT, and PBFT, BECP operates without relying on a global leader, thus guaranteeing greater decentralisation. Unlike PoW, BECP has low resource demands, and unlike PoS, it is resistant to collusion. Simulation and analysis verify that BECP achieves similar performance in throughput compared to classical protocols and offers better scalability, while significantly outperforming epidemic-based approaches like the Snow family in terms of communication overhead and consensus latency. In terms of future work, the BECP consensus protocol can be extended to include the ability to detect node failures and trigger a recovery process to make the system resilient to such failures.

\bibliographystyle{plain}
\bibliography{main.bib}
\end{document}